\documentclass[11pt]{article}
\usepackage{blindtext}
\usepackage{graphicx}
\usepackage{subcaption}
\usepackage{amsmath}
\usepackage{amssymb}
\usepackage{amsthm}
\usepackage{xkeyval}
\title{Squeezing in Both the Plus and Minus Quadratures \\with the Uncertainty Relation Perfectly Holding}
\author{Ayana Yihunie\footnote{Email address: ayana.yihunie@aau.edu.et} and 
Fesseha Kassahun\footnote{Email address: fessehakassahun@gmail.com}\\
                Department of Physics,
                Addis Ababa University\\
                P. O. Box 1176, Addis Ababa, Ethiopia}
\begin{document}
\maketitle
\begin{abstract}
We have considered a cavity mode driven by coherent light and interacting with a three-level atom available in an open cavity 
coupled to a vacuum reservoir. We have carried out our analysis by putting the noise operators associated with  the  vacuum 
reservoir in normal order. We have also considered the interaction of the three-level atom with the  vacuum reservoir outside the cavity. 
It is found that the squeezing occurs in both the plus and the minus quadratures and the maximum quadrature squeezing happens to be 52.08\% and 33.32\% 
below the vacuum state level, respectively. We have established that the uncertainty relation holds perfectly for this case as well. In addition, 
we have found that the squeezing in a pair of superposed cavity modes occurs in the plus and minus quadratures and have the same value. The amount of squeezing 
in each quadrature turns out to be half of the sum of the squeezing in the plus and minus quadratures of each cavity mode. Furthermore, we have 
observed that the presence of spontaneous emission decreases the mean photon number of the cavity mode but does not affect the maximum quadrature squeezing.
\end{abstract}
\hspace*{9.5mm}Keywords: Quadrature squeezing, ~Superposition, ~Mean photon number, Spontaneous\newline
\hspace*{2.9cm}emission
\vspace*{5mm}
\section{Introduction}
\hspace*{2mm} Quadrature squeezing is a nonclassical feature of light and constitutes an interesting subject of quantum optics. 
The quantum properties of squeezed light has been extensively studied by several authors [1-8]. 
In a squeezed state the quantum noise in one quadrature is below the vacuum state level, with the product of the uncertainties in the two quadratures 
satisfying the uncertainty relation. Due to the quantum noise reduction achievable below the vacuum state level, 
squeezed light has potential applications in the detection of  weak signals and in low-noise communications [9]. \\
\hspace*{2mm} There has been a considerable interest to generate squeezed light 
using various quantum optical processes such as subharmonic generation [9, 10 ], second harmonic generation [1, 11], and four-wave mixing [4, 5, 12]. 
Squeezed light can also be generated by a three-level laser under certain conditions [9, 13-19]. 
A three-level laser is a quantum optical system in which light is generated by three-level atoms inside a cavity usually coupled to a vacuum reservoir. 
When a three-level atom makes a transition from the top to bottom level via the intermediate level, two photons are emitted. The two photons are highly 
correlated and this correlation is responsible for the squeezing of the light produced by a three-level laser. Fesseha [15] has studied a three-level 
laser in which three-level atoms available in a closed cavity are pumped to the top level at a constant rate by means of electron bombardment. 
He has found a maximum global quadrature squeezing of 50\% below the vacuum state level. Moreover, 
Fesseha [9] has considered a three-level laser in which the top and bottom levels of three-
level atoms available in a closed cavity are coupled by coherent light.  He has established that the three-level laser under certain conditions 
generates squeezed light with a maximum global quadrature squeezing of 50\% below the vacuum state level. \\
\hspace*{2mm} Here we seek to study the quantum properties of the cavity mode driven by coherent light and interacting with a three-level atom 
available in an open cavity  coupled to a vacuum reservoir via a single-port mirror as shown in Figure 1.  
We carry out our calculation by putting the noise operators associated with the vacuum reservoir in normal order [20] and taking into consideration 
the interaction of the three-level atom with the vacuum reservoir outside the cavity. Thus taking into account 
the damping of the cavity modes by the reservoir and the interaction of the three-level atom with the resonant cavity modes as well as  
the vacuum reservoir outside the cavity, we obtain the quantum Langevin equations for the cavity modes. Moreover, using the master equation 
we derive the equations of evolution of the expectation values of atomic operators. 
Employing the steady-state solutions of the equations of evolutions of the cavity modes and the  atomic operators, we determine the mean 
photon number and the global quadrature squeezing of the cavity  mode. We also calculate the mean photon number 
and the global quadrature squeezing of a pair of superposed cavity  modes.

\section{Operator dynamics}
\hspace*{2mm} We consider here the case in which a three-level atom in a cascade configuration is available in an open cavity driven by coherent light and 
coupled to a vacuum reservoir via a single-port mirror. We denote the top, intermediate, and bottom levels of a three-level atom by 
$ |a\rangle $, $ |b\rangle $, and $ |c\rangle $, respectively. The atom absorbs a photon from the cavity and makes a transition from the bottom level 
to the top level. Then it emits a photon and decays to the intermediate level. Finally, the atom emits another photon and decays to the bottom level. 
Moreover, spontaneous emission takes place from $ |a\rangle $ to  $ |b\rangle $,  $ |b\rangle $ to $ |c\rangle $, 
and $ |a\rangle $ to $ |c\rangle $ with the same emission decay constant. We wish to call the light emitted from the top level cavity mode $a_1$ and 
the one emitted from the intermediate level cavity mode $a_2$. In addition, we represent the cavity mode driven by coherent light by the operator $\hat b$.\\ 
\begin{figure}
\begin{center}
\begin{picture}(250,180)(-180,-100)
\put(-60,55){\linethickness{0.4mm}\line(1,0){70}}\put(10,52.5){$|a\rangle$}
\put(-3,55){\linethickness{0.4mm}\vector(0,-1){40}}\put(-1,27.5){$\omega_{a}$}
\put(-40,55){\linethickness{0.4mm}\vector(0,-1){100}}\put(-40,27.5){$\gamma$}
\put(-19,55){\linethickness{0.4mm}\vector(0,-1){40}}\put(-18,27.5){$\gamma_c$}
\put(-30,55){\linethickness{0.4mm}\vector(0,-1){40}}\put(-30,27.5){$\gamma$}
\put(-60,15){\linethickness{0.4mm}\line(1,0){70}}\put(10,12.5){$|b\rangle$}
\put(-19,15){\linethickness{0.4mm}\vector(0,-1){60}}\put(-18,-22.5){$\gamma_c$}
\put(-30,15){\linethickness{0.4mm}\vector(0,-1){60}}\put(-30,-22.5){$\gamma$}
\put(-3,15){\linethickness{0.4mm}\vector(0,-1){60}}\put(-1,-22.5){$\omega_{b}$}
\put(-60,-45){\linethickness{0.4mm}\line(1,0){70}}\put(10,-47.5){$|c\rangle$}
\put(-52,-45){\linethickness{0.4mm}\vector(0,1){100}}
\put(-120,-60){\linethickness{1mm}\line(0,1){135}}
\put(-125,-60){\linethickness{0.4mm}\line(0,1){135}}
\put(-180.5,10){\linethickness{0.4mm}\vector(1,0){55}}\put(-154,14.5){$\bf{\eta}$}
\put(-120,10){\linethickness{0.4mm}\vector(1,0){40}}\put(-105,14.5){$\bf{\hat b}$}
\put(40,-60){\linethickness{0.58mm}\line(0,1){135}}
\put(44,-60){\linethickness{0.5mm}\line(0,1){135}}
\put(44,10){\linethickness{0.4mm}\vector(1,0){60}}\put(72,14.5){$\bf{\kappa}$}
\end{picture}
\caption{\footnotesize{Schematic representation of a three-level atom in a cascade configuration available in an open 
cavity driven by coherent light and coupled to vacuum reservoir via a single-port mirror}.}
\end{center}
\label{1}
\end{figure}
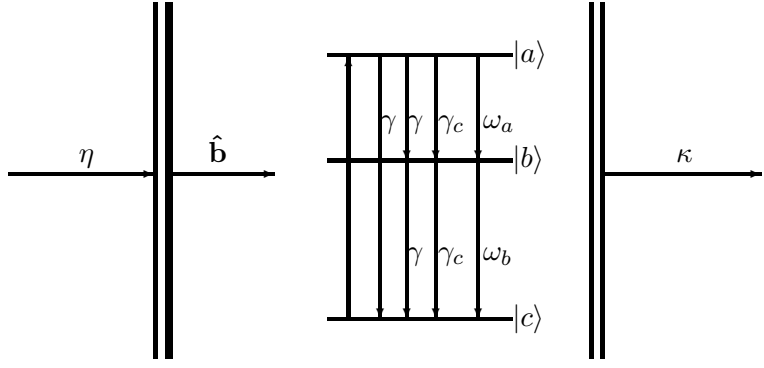
\hspace*{2mm}  The interaction of the driving coherent light with the resonant cavity mode can be  described by the Hamiltonian
\begin{equation}\label{8172}
\hat H'=i\lambda(\hat c\hat b^\dagger-\hat b\hat c^\dagger),
\end{equation}
where $\hat c$ and $\hat b$ are  
the annihilation  operators for the driving coherent light and the cavity mode, respectively  and $\lambda$ is the coupling constant. 
In order to have a mathematically manageable analysis, we replace the operator $\hat c$ by a real and constant c-number $ \mu$. 
On account of this, we can rewrite Eq. \eqref{8172} as 
\begin{equation}\label{8173}
\hat H'=i\eta(\hat b^\dagger-\hat b),
\end{equation}
where $\eta=\lambda\mu$.
In addition, the interaction of a three-level atom with the resonant cavity modes $b$, $a_1$, and $a_2$ can be described by the Hamiltonian
\begin{equation}\label{8174}
\hat H^{''}=ig\left(\hat \sigma_c^{\dagger }\hat b-\hat b^\dagger\hat\sigma_c +\hat\sigma_a^{\dagger }\hat a_1-\hat a_1^\dagger\hat \sigma_a+
\hat\sigma_b^{\dagger} \hat a_2 -\hat a_2^\dagger \hat \sigma_b \right),
\end{equation}
where
\begin{equation}\label{8175}
\hat \sigma_a=|b\rangle\langle a|,
\end{equation}
\begin{equation}\label{8176}
\hat \sigma_b=|c\rangle\langle b|,
\end{equation}
\begin{equation}\label{8177}
 \hat \sigma_c=|c\rangle\langle a|,
\end{equation}
are lowering atomic operators,  $\hat a_{1} (\hat a_{2})$ is the annihilation operator for the cavity mode $a_1$ ($a_2$), with $g$ being  
the coupling constant between the atom and the cavity mode $ b$ or $ a_1$ or $ a_2$.
Hence  taking into account  Eqs. \eqref{8173} and \eqref{8174}, the total Hamiltonian of the system under consideration can be put in the form
\begin{eqnarray}\label{8178}
\hat H= i\eta(\hat b^\dagger-\hat b)+ig\left(\hat \sigma_c^{\dagger }\hat b-\hat b^\dagger\hat\sigma_c +\hat\sigma_a^{\dagger }\hat a_1-\hat a_1^\dagger\hat \sigma_a+
\hat\sigma_b^{\dagger} \hat a_2 -\hat a_2^\dagger \hat \sigma_b \right).
\end{eqnarray}
\hspace*{2mm} We carry out our calculation by putting the noise operators associated with the vacuum reservoir in normal order. 
Thus the noise operators will not have any effect on the dynamics of the cavity mode operators. 
We can therefore drop the noise operators and write the quantum Langevin equations for the operators $\hat{a_1}$, $\hat{a_2}$, and $\hat{b}$ as
\begin{equation}\label{8179}
{d\hat{a_1}\over dt}=-{\kappa\over 2}\hat{a_1}-i[\hat{a_1},\hat{H}],
\end{equation}
\begin{equation}\label{8180}
{d\hat{a_2}\over dt}=-{\kappa\over 2}\hat{a_2}-i[\hat{a_2},\hat{H}],
\end{equation}
\begin{equation}\label{8181}
{d\hat{b}\over dt}=-{\kappa\over 2}\hat{b}-i[\hat{b},\hat{H}],
\end{equation}
where $\kappa$ is the cavity damping constant. Then with the aid of Eq. \eqref{8178} and following the discussion given in  Ref [21],  we easily find
\begin{equation}\label{8182}
{d\hat{a_1}\over dt}=-{\kappa\over 2}\hat{a_1}-g\hat{\sigma}_{a}
\end{equation}
\begin{equation}\label{8183}
{d\hat{a_2}\over dt}=-{\kappa\over 2}\hat{a_2}-g\hat{\sigma}_{b}.
\end{equation}
\begin{equation}\label{8184}
{d\hat{b}\over dt}=-{\kappa\over 2}\hat{b}+\eta- g\hat {\sigma}_{c}.
\end{equation}
\hspace*{2mm} On the other hand, the  master equation for a three-level atom interacting with the cavity modes and the vacuum reservoir outside 
the cavity can be written as [9]
\begin{eqnarray}\label{8185}
\frac{d\hat\rho}{dt}&=&-i[\hat H'', \hat \rho]+\frac{\gamma}{2}\Big(2\sigma_a\hat\rho\sigma_a^{\dagger } -\sigma_a^{\dagger }\sigma_a\hat\rho 
  -\hat\rho\sigma_a^{\dagger }\sigma_a\Big)
  + \frac{\gamma}{2}\Big(2\sigma_b\hat\rho\sigma_b^{\dagger } -\sigma_b^{\dagger }\sigma_b\hat\rho
  - \hat\rho\sigma_b^{\dagger }\sigma_b\Big)\nonumber\\
 &&+ \frac{\gamma}{2}\Big(2\sigma_c\hat\rho\sigma_c^{\dagger } -\sigma_c^{\dagger }\sigma_c\hat\rho-\hat\rho\sigma_c^{\dagger }\sigma_c\Big),
\end{eqnarray}
where 
$\gamma$ is the spontaneous emission decay constant. Now with the aid of Eq. \eqref{8174}, one can put  \eqref{8185}  in the form
\begin{eqnarray}\label{8186}
\frac{d\hat\rho(t)}{dt}&=&g\Big(\hat \sigma_c^{\dagger }\hat b\hat\rho-\hat b^\dagger\hat\sigma_c\hat\rho
+\hat\sigma_a^{\dagger }\hat a_1\hat\rho-\hat a_1^\dagger\hat \sigma_a\hat\rho
+\hat\sigma_b^{\dagger }\hat a_2\hat\rho 
-\hat a_2^\dagger\hat \sigma_b\hat\rho
-\hat\rho\hat \sigma_c^{\dagger }\hat b\nonumber\\
&&+\hat\rho\hat b^\dagger\hat\sigma_c-\hat\rho\hat\sigma_a^{\dagger }\hat a_1
+\hat\rho\hat a_1^\dagger\hat \sigma_a
-\hat\rho\hat\sigma_b^{\dagger }\hat a_2+\hat\rho\hat a_2^\dagger\hat \sigma_b\Big)\nonumber\\
&&+\frac{\gamma}{2}\Big(2\hat\sigma_a\hat\rho\hat\sigma_a^{\dagger } -\hat\sigma_a^{\dagger }\hat\sigma_a\hat\rho 
  -\hat\rho\hat\sigma_a^{\dagger }\hat\sigma_a\Big)
  + \frac{\gamma}{2}\Big(2\hat\sigma_b\hat\rho\hat\sigma_b^{\dagger } -\hat\sigma_b^{\dagger }\hat\sigma_b\hat\rho-
 \hat\rho\hat\sigma_b^{\dagger }\hat\sigma_b\Big)\nonumber\\
 &&+\frac{\gamma}{2}\Big(2\sigma_c\hat\rho\sigma_c^{\dagger } -\sigma_c^{\dagger }\sigma_c\hat\rho-\hat\rho\sigma_c^{\dagger }\sigma_c\Big).
\end{eqnarray}
Furthermore, applying the relation
\begin{equation}\label{8187}
 \frac{d}{dt}\langle\hat A\rangle=Tr\left(\frac{d\hat\rho(t)}{dt}\hat A\right) 
\end{equation}
along with Eq. \eqref{8186} and the cyclic property of the trace operation, one can readily establish that
\begin{equation}\label{8188}
\frac{d}{dt}\langle\hat \sigma_a\rangle=-\frac{3}{2}\gamma \langle\hat\sigma_a\rangle+ g\langle\hat\sigma_b^{\dagger } \hat b\rangle+g \langle (\hat\eta_b - \hat\eta_a)\hat a_1\rangle
+g\langle \hat a_2^\dagger\hat\sigma_c\rangle,
\end{equation}
\begin{equation}\label{8189}
\frac{d}{dt}\langle\hat \sigma_b\rangle=-\frac{1}{2}\gamma\langle\hat\sigma_b\rangle-g\langle\hat a_1^\dagger\hat\sigma_c \rangle+g\langle (\hat\eta_c- \hat\eta_b)\hat a_2\rangle
-g\langle \hat\sigma_a^{\dagger }\hat b\rangle,
\end{equation}
\begin{equation}\label{8190}
\frac{d}{dt}\langle\hat \sigma_c\rangle= -\gamma\langle\hat\sigma_c\rangle+g\langle\hat\sigma_b\hat a_1 \rangle+ g\langle (\hat\eta_c - \hat\eta_a)\hat b\rangle
-g\langle \hat\sigma_a\hat a_2\rangle,
\end{equation}
\begin{equation}\label{8191}
\frac{d}{dt}\langle\hat \eta_a\rangle=-2\gamma\langle\hat\eta_a\rangle+ g\langle\hat\sigma_c^{ \dagger }\hat b \rangle+ g\langle\hat\sigma_a^{ \dagger } \hat a_1\rangle
+g\langle \hat b^\dagger\hat\sigma_c\rangle+g \langle\hat a_1^\dagger\hat\sigma_a \rangle,
\end{equation}
\begin{equation}\label{8192}
\frac{d}{dt}\langle\hat \eta_b\rangle= -\gamma\langle\hat\eta_b\rangle+\gamma\langle\hat\eta_a\rangle+ g\langle\hat\sigma_b^{ \dagger }\hat a_2 \rangle - g\langle\hat\sigma_a^{ \dagger } \hat a_1\rangle
+g\langle \hat a_2^\dagger\hat\sigma_b\rangle-g\langle\hat a_1^\dagger\hat\sigma_a \rangle,
\end{equation}
where 
\begin{equation}\label{8193}
 \hat\eta_a=|a\rangle\langle a|,
\end{equation}
\begin{equation}\label{8194}
 \hat\eta_b=|b\rangle\langle b|,
\end{equation}
We see that Eqs. \eqref{8188}-\eqref{8192}, are nonlinear differential equations and hence it is not possible to obtain the exact 
time dependent solutions of these equations. We intend to overcome this problem by applying the large-time approximation [20]. 
Then using this approximation scheme, we get from Eqs. \eqref{8182}, \eqref{8183}, and \eqref{8184}  the approximately valid relations
\begin{equation}\label{8195}
  \hat a_1= -\frac{2g}{\kappa}\hat \sigma_a,
\end{equation}
\begin{equation}\label{8196}
 \hat a_2= -\frac{2g}{\kappa}\hat \sigma_b,
\end{equation}
\begin{equation}\label{8197}
  \hat b= \frac{2\eta}{\kappa} -\frac{2g}{\kappa}\hat \sigma_c.
\end{equation}
Evidently, these would turn out to be exact relations at steady state.
Introducing ~Eqs. \eqref{8195}, \eqref{8196}, and (\ref{8197}) into Eqs. \eqref{8188}-\eqref{8192}, we get
\begin{equation}\label{8198}
\frac{d}{dt}\langle\hat \sigma_a\rangle= -\frac{3}{2}(\gamma_c+\gamma)\langle\hat\sigma_a\rangle+\varepsilon\langle\hat\sigma_b^{\dagger}\rangle,
\end{equation}
\begin{equation}\label{8199}
\frac{d}{dt}\langle\hat \sigma_b\rangle=-\frac{1}{2}(\gamma_c+\gamma)\langle\hat\sigma_b \rangle -\varepsilon\langle \hat\sigma_a^{\dagger}\rangle,
\end{equation}
\begin{equation}\label{8200}
\frac{d}{dt}\langle\hat \sigma_c\rangle= -(\gamma_c+\gamma)\langle \hat\sigma_c\rangle+\varepsilon(\langle \hat\eta_c \rangle - \langle\hat\eta_a\rangle),
\end{equation}
\begin{equation}\label{8201}
\frac{d}{dt}\langle\hat \eta_a\rangle= -2(\gamma_c+\gamma)\langle\hat\eta_a\rangle+\varepsilon(\langle\hat\sigma_c^{ \dagger}\rangle+ \langle\hat\sigma_c\rangle),
\end{equation}
\begin{equation}\label{8202}
\frac{d}{dt}\langle\hat \eta_b\rangle = -(\gamma_c+\gamma)\langle\hat\eta_b\rangle+(\gamma_c+\gamma)\langle\hat\eta_a\rangle,
\end{equation}
where
\begin{equation}\label{8203}
 \gamma_c=\frac{4g^2}{\kappa}
\end{equation}
is the stimulated emission decay constant and $\varepsilon$ is defined by
\begin{equation}\label{8204}
 \varepsilon=\frac{2g\eta}{\kappa}.
 \end{equation}
\hspace*{2mm} We note that the steady-state solutions of Eqs. \eqref{8198}-\eqref{8202} are given by
\begin{equation}\label{8205}
 \langle \hat \sigma_a\rangle=\frac{2\varepsilon}{3(\gamma_c+\gamma)}\langle\hat\sigma^{\dagger}_b\rangle,
\end{equation}
\begin{equation}\label{8206}
 \langle \hat \sigma_b\rangle=-\frac{2\varepsilon}{(\gamma_c+\gamma)}\langle\hat\sigma^{\dagger}_a\rangle,
\end{equation}
\begin{equation}\label{8207}
 \langle \hat \sigma_c\rangle=\frac{\varepsilon}{(\gamma_c+\gamma)}(\langle\hat \eta_c\rangle-\langle\hat \eta_a\rangle),
\end{equation}
\begin{equation}\label{8208}
 \langle \hat \eta_a\rangle=\frac{\varepsilon}{2(\gamma_c+\gamma)}(\langle\hat\sigma^{\dagger}_c\rangle+\langle\hat \sigma_c\rangle),
\end{equation}
\begin{equation}\label{8209}
 \langle \hat \eta_b\rangle=\langle \hat \eta_a\rangle.
\end{equation}
Applying Eqs. \eqref{8205} and \eqref{8206}, we easily find 
\begin{equation}\label{8210}
 \langle \hat \sigma_a\rangle=\langle \hat \sigma_b\rangle=0.
\end{equation}
Furthermore, with the aid of the identity 
\begin{equation}\label{8211}
  \hat\eta_a + \hat\eta_b+ \hat\eta_c= \hat I,
\end{equation}
we see that 
\begin{equation}\label{8212}
 \langle \hat\eta_a\rangle+\langle \hat\eta_b\rangle+\langle\hat\eta_c\rangle=1.
\end{equation}
We interpret $\langle \hat\eta_a\rangle$, $\langle \hat\eta_b\rangle$, and $\langle\hat\eta_c\rangle$ as the probabilities for 
the three-level atom to be in the upper, intermediate, and bottom levels, respectively.
Now using Eq. \eqref{8212} along with Eqs. \eqref{8207}, \eqref{8208}, and \eqref{8209}, we easily find 
\begin{equation}\label{8213}
 \langle \hat \sigma_c\rangle=\frac{\varepsilon(\gamma_c+\gamma)}{(\gamma_c+\gamma)^2+3\varepsilon^2},
\end{equation}
\begin{equation}\label{8214}
 \langle \hat \eta_a\rangle=\langle \hat \eta_b\rangle=\frac{\varepsilon^2 }{(\gamma_c+\gamma)^2+3\varepsilon^2},
\end{equation}
\begin{equation}\label{8215}
 \langle \hat \eta_c\rangle=\frac{\varepsilon^2+(\gamma_c+\gamma)^2}{(\gamma_c+\gamma)^2+3\varepsilon^2}.
\end{equation}
\section{Cavity mode $b$}
\hspace*{2mm} Here we wish to calculate the mean photon number and the quadrature squeezing of the cavity mode $b$ at steady state. 
\subsection{The mean photon number}
\hspace*{2mm} The mean photon number for the cavity mode $b$ at steady state is defined by $\bar n=\langle\hat b^{\dagger}\hat b\rangle$. 
Then applying  Eq. \eqref{8197}, we easily find 
\begin{figure}
\centering
 \includegraphics[scale=0.8]{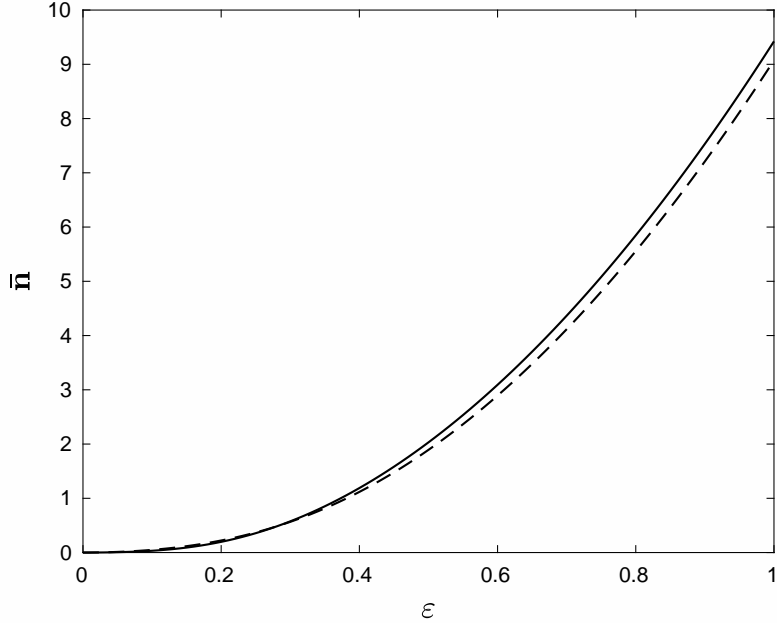}
  \caption{\footnotesize{  Plots of Eq. \eqref{8217} for 
  $\gamma_c=0.5 $, $\kappa=0.8$, and for $\gamma=0$ (solid curve), and for $\gamma=0.3$ (dashed curve)}.}
 \label{2}
\end{figure}
\begin{equation}\label{8216}
\bar n= \frac{4\eta^2}{\kappa^2}-\frac{8\eta g}{\kappa^2}\langle\hat \sigma_c\rangle+ \frac{\gamma_c}{\kappa}\langle \hat \eta_a\rangle.
\end{equation}
The first term represents the mean photon number for the cavity mode b due to the
driving coherent light, the second term represents the mean number of photons absorbed by the atom, and the third term represents the mean number 
of photons emission by the three-level atom. Now on account of Eqs. \eqref{8213} and \eqref{8214} along with \eqref{8204}, we obtain 
\begin{equation}\label{8217}
 \bar n= \frac{4\varepsilon^2}{\kappa\gamma_c}-\frac{\varepsilon^2}{\kappa}\left[\frac{(3\gamma_c+4\gamma)}{(\gamma_c+\gamma)^2+3\varepsilon^2}\right].
\end{equation}
In the absence of spontaneous emission ($\gamma = 0$) the mean photon number for the cavity mode $b$ takes the form
\begin{equation}\label{8218}
 \bar n= \frac{4\varepsilon^2}{\kappa\gamma_c}-\frac{\varepsilon^2}{\kappa}\left[\frac{3\gamma_c}{\gamma_c^2+3\varepsilon^2}\right].
\end{equation}
The plots in Figure \ref{2} show that the presence of spontaneous emission decreases the mean photon number of the cavity light.
\subsection{Quadrature squeezing}
\hspace*{2mm} We next wish to calculate the quadrature squeezing of the cavity mode. To this end, the squeezing properties of 
the cavity mode are described by two quadrature operators defined by 
\begin{equation}\label{8219}
\hat{b}_{+}=\hat{b}^{\dag}+\hat{b}
\end{equation}
and
\begin{equation}\label{8220}
\hat{b}_{-}=i(\hat{b}^{\dag}-\hat{b}).
\end{equation}
It can be readily established that
\begin{equation}\label{8221}
  [\hat b_{-}, \hat b_{+}]= 2i\frac{\gamma_c}{\kappa}(\hat \eta_a-\hat \eta_c).
 \end{equation}
The uncertainty relation for two physical observables $ A$ and $B$, satisfying  
the commutation relation [22]
\begin{equation}\label{8222}
 [\hat A, \hat B]=i \hat C,
\end{equation}
has the form 
\begin{equation}\label{8223}
 \Delta A \Delta B \geq {1\over2}\left|\langle \hat C\rangle\right|. 
\end{equation}
It then follows that  
\begin{equation}\label{8224}
 \Delta b_{+} \Delta b_{-} \geq \frac{\gamma_c}{\kappa} \left|\langle\hat \eta_a\rangle -\langle\hat \eta_c\rangle\right|.
\end{equation}
Upon substituting  Eqs. \eqref{8214} and \eqref{8215} into  Eq. \eqref{8224}, we have
 \begin{equation}\label{8225}
 \Delta b_{+} \Delta b_{-} \geq f_{a}(\varepsilon),
 \end{equation}
  where
  \begin{equation}\label{8226}
f_{a}(\varepsilon)=\frac{\gamma_c }{\kappa}\left(\frac{(\gamma_c+\gamma)^2}{3\varepsilon^2+(\gamma_c+\gamma)^2}\right).
\end{equation}
On setting $\varepsilon=0$ in Eq. \eqref{8225}, we see that
\begin{equation}\label{8227}
 \Delta b_{+} \Delta b_{-} \geq \frac{\gamma_c}{\kappa}.
\end{equation}
This represents the quadrature uncertainty relation for a vacuum state.\\
\hspace*{2mm} The variances of the plus and the minus quadrature operators are expressible as 
 \begin{equation}\label{8228}
 (\Delta b_{+})^2=\langle \hat b^{\dagger}\hat b\rangle
 +\langle \hat b\hat b^{\dagger}\rangle + \langle \hat b^{\dagger 2}\rangle + \langle \hat b^{2}\rangle -\langle\hat b^{\dagger}\rangle^2-\langle\hat b\rangle^2-2\langle\hat b^\dagger\rangle\langle\hat b\rangle 
\end{equation}
and
 \begin{equation}\label{8229}
 (\Delta b_{-})^2=\langle \hat b^{\dagger}\hat b\rangle
 +\langle \hat b\hat b^{\dagger}\rangle -\langle \hat b^{\dagger 2}\rangle -\langle \hat b^{2}\rangle +\langle\hat b^{\dagger}\rangle^2
 +\langle\hat b\rangle^2-2\langle\hat b^\dagger\rangle\langle\hat b\rangle.
\end{equation}
Now employing  Eq. \eqref{8197}, one can easily verify that 
\begin{equation}\label{8230}
  \langle\hat b^2\rangle=\frac{4\varepsilon^2}{\kappa\gamma_c}-\frac{4\varepsilon}{\kappa}\langle\hat \sigma_c\rangle,
 \end{equation}
\begin{equation}\label{8231}
 \langle\hat b\rangle^2=\frac{4\varepsilon^2}{\kappa\gamma_c}
 -\frac{4\varepsilon }{\kappa}\langle\hat \sigma_c\rangle+\frac{\gamma_c}{\kappa}\langle\hat \sigma_c\rangle^2,
\end{equation}
 \begin{equation}\label{8232}
  \langle\hat b^\dagger\rangle\langle\hat b\rangle=\frac{4\varepsilon^2}{\kappa\gamma_c}
  -\frac{4\varepsilon }{\kappa}\langle\hat \sigma_c\rangle+\frac{\gamma_c}{\kappa}\langle\hat \sigma_c\rangle^2,
 \end{equation}
 \begin{equation}\label{8233}
 \langle\hat b \hat b^\dagger\rangle=\frac{4\varepsilon^2}{\kappa\gamma_c}
 -\frac{4\varepsilon }{\kappa}\langle\hat \sigma_c\rangle+\frac{\gamma_c}{\kappa}\langle\hat \eta_c\rangle.
\end{equation}
\begin{figure}
\centering
 \includegraphics[scale=0.8]{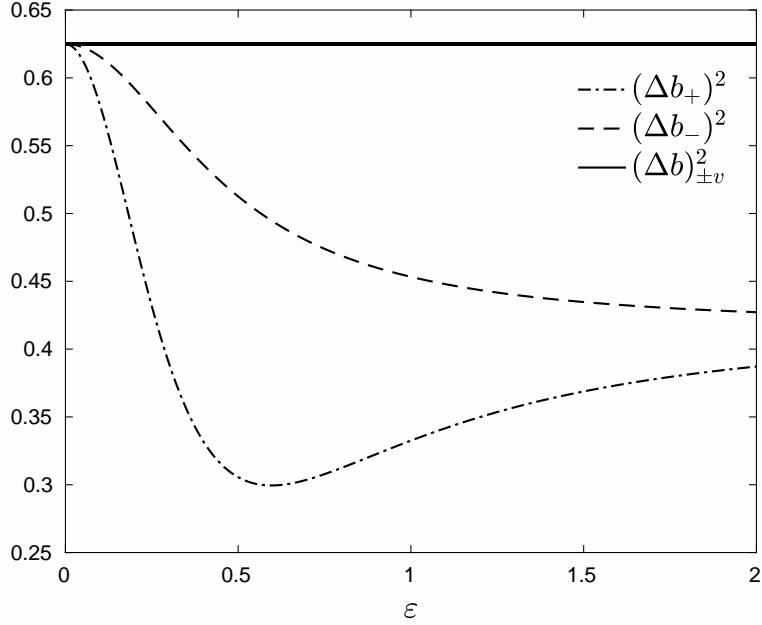}
  \caption{\footnotesize{ Plots of $(\Delta b_{-})^2$, $(\Delta b_{+})^2$, and $((\Delta b)_{+v}^2=(\Delta b)_{-v}^2)$  versus $\varepsilon$  
  for $\gamma_c=0.5$, $\gamma=0.3$, and   $\kappa=0.8$ }.}
 \label{3}
\end{figure}
On account of Eqs. \eqref{8216}, \eqref{8230}, \eqref{8231}, \eqref{8232}, and \eqref{8233}, we readily obtain 
\begin{equation}\label{8234}
  (\Delta b_{+})^2=\frac{\gamma_c}{\kappa}[\langle\hat \eta_a\rangle+\langle\hat \eta_c\rangle-4\langle\hat \sigma_c\rangle^2],
\end{equation}
\begin{equation}\label{8235}
 (\Delta b_{-})^2=\frac{\gamma_c}{\kappa}(\langle\hat \eta_a\rangle+\langle\hat \eta_c\rangle).
\end{equation}
Therefore, in view of Eqs. \eqref{8213}, \eqref{8214}, and \eqref{8215},  we get 
\begin{equation}\label{8236}
 (\Delta b_{+})^2=\frac{\gamma_{c}}{\kappa}\left[\frac{6\varepsilon^4+(\gamma_c+\gamma)^2((\gamma_c+\gamma)^2+\varepsilon^2)}{((\gamma_c+\gamma)^2+3\varepsilon^2)^2}\right],
  \end{equation}
\begin{equation}\label{8237}
 (\Delta b_{-})^2=\frac{\gamma_{c}}{\kappa}\left[\frac{2\varepsilon^2+(\gamma_c+\gamma)^2}{3\varepsilon^2+(\gamma_c+\gamma)^2}\right]. 
\end{equation}
Upon setting $\varepsilon=0$, in Eqs. \eqref{8236} and \eqref{8237}, we have
 \begin{equation}\label{8238}
  (\Delta b_{+})_v^2= (\Delta b_{-})_v^2=\frac{\gamma_c}{\kappa}.
 \end{equation}
This indeed represents the quadrature variance of the cavity vacuum state in which the uncertainties in the two quadratures are equal and 
satisfy the minimum uncertainty relation given by Eq. \eqref{8227}. From the plots in Figure \ref{3}, 
we observe that both $(\Delta b_{+})^2$ and $(\Delta b_{-})^2$ are below the quadrature 
variance of the vacuum state. This is just unexpected result. 
\begin{figure}
\centering
 \includegraphics[scale=0.8]{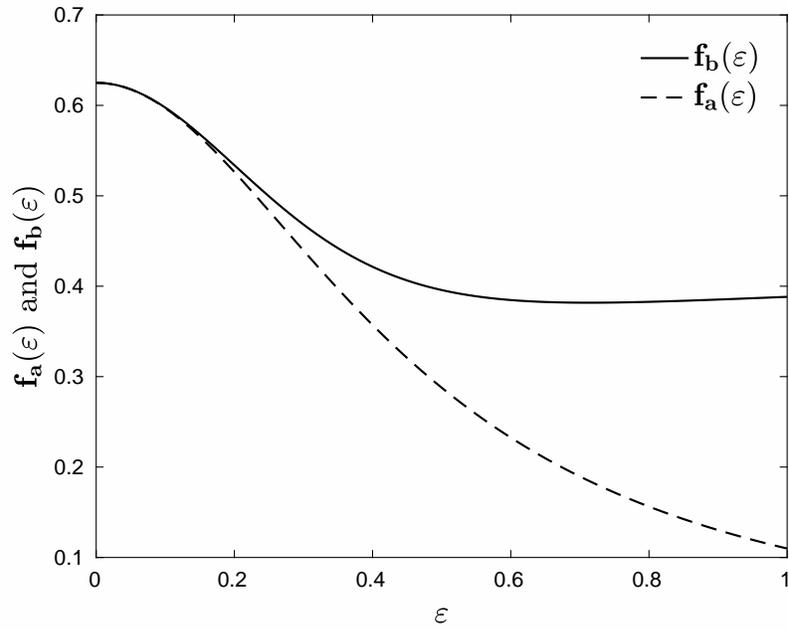}
  \caption{\footnotesize{ Plots of $f_a(\varepsilon)$  and $f_b(\varepsilon)$ versus $\varepsilon$  for $\gamma_c=0.5$, $\gamma=0.3$, and 
  $\kappa=0.8$}.}
 \label{4}
\end{figure}
We then see that the cavity mode $b$ is in a squeezed state in the plus as well as in the minus quadrature. With the aid of 
Eqs. \eqref{8236} and \eqref{8237}, one can write  
\begin{equation}\label{8239}
 \Delta b_{+}\Delta b_{-}=f_b(\varepsilon),
\end{equation}
where
\begin{equation}\label{8240}
 f_b(\varepsilon)=\frac{\gamma_{c}}{\kappa}
 \sqrt{\frac{(6\varepsilon^4+(\gamma_c+\gamma)^2((\gamma_c+\gamma)^2+\varepsilon^2))((\gamma_c+\gamma)^2+2\varepsilon^2)}{((\gamma_c+\gamma)^2+3\varepsilon^2)^3}}.
\end{equation}
We note from the plots in 
Figure \ref{4} that the plots for $f_b({\varepsilon})$ (solid curve) and $f_a({\varepsilon})$ (dashed curve) are equal for 
 $0<\varepsilon\leq 0.03$. However, 
the plot for $f_b({\varepsilon})$ is greater than the plot for $f_a({\varepsilon})$ for $\varepsilon> 0.03$. 

\begin{figure}
\centering
 \includegraphics[scale=0.8]{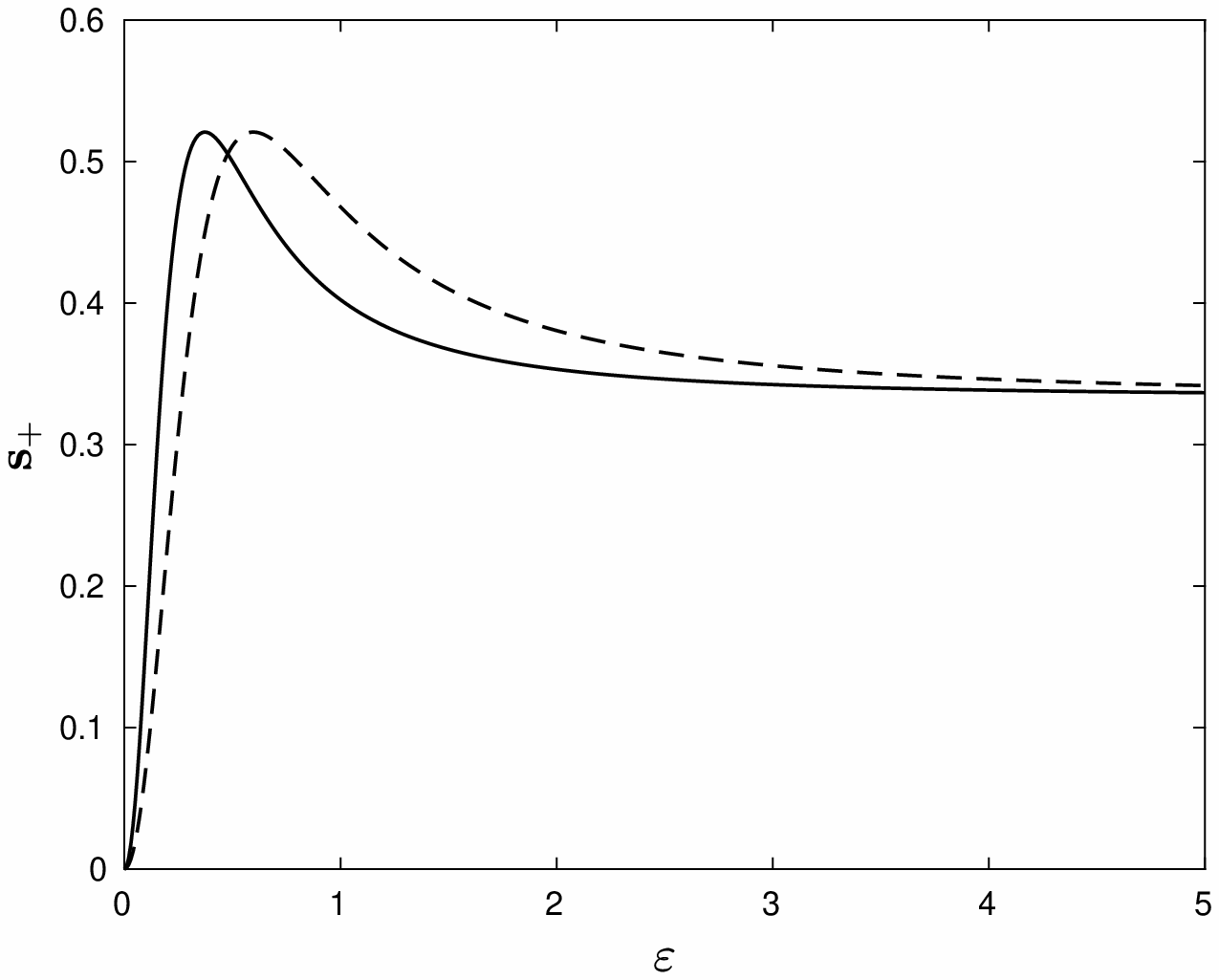}
  \caption{\footnotesize{ Plots of $S_{+}$ versus $\varepsilon$  for $\gamma_c=0.5$, $\kappa=0.8$ , and for $\gamma=0.3$ (dashed curve), and or  $\gamma=0$ (solid curve)
    }.}
 \label{5}
\end{figure}
\begin{figure}
\centering
 \includegraphics[scale=0.8]{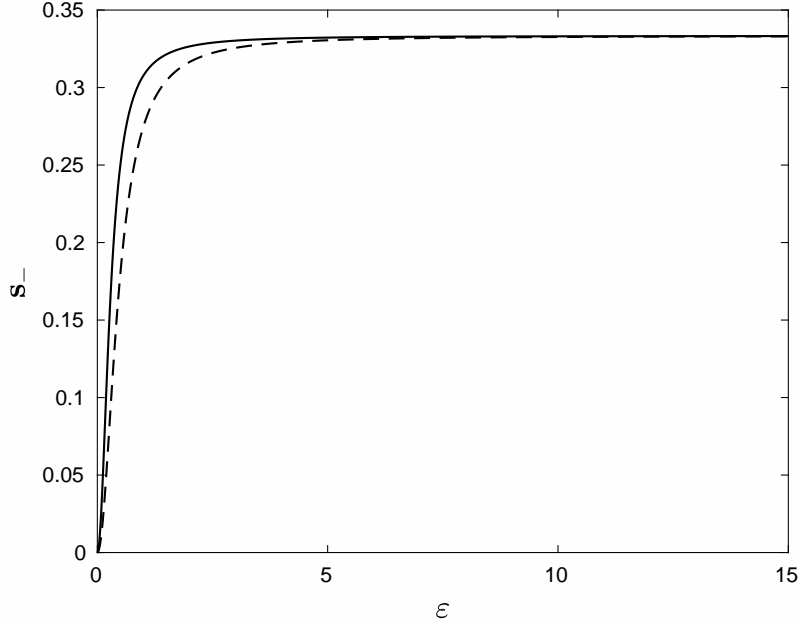}
  \caption{\footnotesize{ Plots of $S_{-}$ versus $\varepsilon$  for $\gamma_c=0.5$, $\kappa=0.8$, and for  $\gamma=0.3$ (dashed curve), and or $\gamma=0$ (solid curve)}.}
 \label{6}
\end{figure}
On the basis of these results, we realize that  the cavity mode $b$ satisfies the uncertainty relation.
\hspace*{2mm} The squeezing of the plus and the minus quadratures of the cavity mode $b$ are defined relative to the quadrature
variance of the vacuum state as [9]
\begin{equation}\label{8241}
 S_+=\frac{(\Delta b_{+})_{v}^2-(\Delta b_{+})^2}{(\Delta b_{+})_{v}^2},
\end{equation}
\begin{equation}\label{8242}
 S_-=\frac{(\Delta b_{-})_{v}^2-(\Delta b_{-})^2}{(\Delta b_{-})_{v}^2}.
\end{equation}
Hence in view of \eqref{8236}, \eqref{8237}, and \eqref{8238}, one can put Eqs. \eqref{8241} and \eqref{8242} in the form 
\begin{equation}\label{8243}
 S_+=1-\left[\frac{6\varepsilon^4+(\gamma_c+\gamma)^2((\gamma_c+\gamma)^2+\varepsilon^2)}{((\gamma_c+\gamma)^2+3\varepsilon^2)^2}\right],
\end{equation}
\begin{equation}\label{8244}
 S_-=1-\left[\frac{2\varepsilon^2+(\gamma_c+\gamma)^2}{3\varepsilon^2+(\gamma_c+\gamma)^2}\right].
\end{equation}
The maximum  squeezing for the plus quadrature (Figure \ref{5}) is $52.08\%$ in the presence (dashed curve) or absence (solid curve) 
of spontaneous emission and occurs at $\varepsilon=0.6$ or $\varepsilon=0.37$, respectively. On the other hand, the squeezing 
of the plus quadrature in the presence of spontaneous emission is less than the squeezing of the plus quadrature in the absence of spontaneous emission 
for  $0<\varepsilon\leq0.48$. But the squeezing of the plus quadrature in the absence of spontaneous emission is less than 
that of the squeezing of the plus quadrature in the presence of spontaneous emission for  $\varepsilon>0.48$. In addition, we see from the plots 
in  Figure \ref{6} that the maximum squeezing for the minus quadrature is $33.32\%$ in the presence or absence of spontaneous emission and occurs 
for $\varepsilon\geq 15$.
\section{Superposed cavity modes}
\hspace*{2mm} In this section we seek to calculate the mean photon number and the global quadrature squeezing of a pair of superposed cavity modes each 
driven by coherent light and interacting 
with a three-level atom. To this end, we wish to represent the two cavity  modes by the annihilation operators $\hat a$ and $\hat b$. We 
represent the superposed cavity modes by the operator $\hat c$. We then obtain the equation of evolution for the superposed cavity modes. Applying the 
steady-state solution of the resulting equation, we calculate the mean photon number and the global quadrature squeezing.\\
\hspace*{2mm} According to Fesseha [21], we can define the operator representing the superposition of the cavity  modes by 
\begin{equation}\label{8245}
 \hat c=\hat a+i\hat b.
\end{equation}
Applying this equation, one can write 
\begin{equation}\label{8246}
 \frac{d\hat c}{dt}=\frac{d\hat a}{dt}+i\frac{d\hat b}{dt}.
\end{equation}
On the basis of Eq. \eqref{8184}, one can write the equation of evolution for the cavity  modes $ a$ and $ b$ as 
\begin{equation}\label{8247}
 \frac{d \hat a}{dt}= -\frac{\kappa}{2}\hat a + \eta- g\hat \sigma_c
\end{equation}
and 
\begin{equation}\label{8248}
 \frac{d \hat b}{dt}= -\frac{\kappa}{2}\hat b + \eta- g\hat \sigma_c'.
\end{equation}
there follows 
\begin{equation}\label{8249}
 \frac{d \hat c}{dt}= -\frac{\kappa}{2}(\hat a+i\hat b) + \eta(1+i)- g(\hat \sigma_c+i\hat \sigma_c').
\end{equation}
In view of Eq. \eqref{8245}, we rewrite Eq. \eqref{8249} as 
\begin{equation}\label{8250}
 \frac{d \hat c}{dt}= -\frac{\kappa}{2}\hat c + \eta(1+i)- g\hat \sigma_s,
\end{equation}
where
\begin{equation}\label{8251}
\hat \sigma_s=\hat \sigma_c+i\hat \sigma_c',
\end{equation}
with 
\begin{equation}\label{8252}
\hat\sigma_c=|c\rangle\langle a|,
\end{equation}
\begin{equation}\label{8253}
\hat\sigma_c'=|c'\rangle\langle a'|
\end{equation}
are  lowering atomic operators and $s$ stands for superposition. The steady-state solution of Eq. \eqref{8250} is 
\begin{equation}\label{8254}
 \hat c=\frac{2\eta}{\kappa}(1+i)-\frac{2g}{\kappa}\hat \sigma_s.
\end{equation}
In addition, with the aid of Eq. \eqref{8251}, one can readily verify that 
\begin{equation}\label{8255}
\hat \sigma_s^\dagger\hat \sigma_s=\hat \eta_a+\hat \eta'_a,
\end{equation}
\begin{equation}\label{8256}
\hat \sigma_s\hat \sigma_s^\dagger=\hat \eta_c+\hat \eta'_c,
\end{equation}
\begin{equation}\label{8257}
\hat \sigma_s^2=0,
\end{equation}
where  $\hat \eta_a$ and $\hat \eta'_a$ representing the probability for  atoms to be in the upper level, and $\hat \eta_c$ and
$\hat \eta'_c$ representing the probability for the  atoms to be in the bottom level.
\subsection{The mean photon number}
\hspace*{2mm} The mean photon number for the superposed cavity  modes is given by 
\begin{equation}\label{8258}
 \bar n_s=\langle \hat c^\dagger\hat c\rangle.
\end{equation}
Hence using Eq. \eqref{8254}, we easily find 
\begin{equation}\label{8259}
 \bar n_s=\frac{8\eta^2}{\kappa^2}-\frac{4\eta g}{\kappa^2}[(1-i)\langle\hat\sigma_s\rangle+(1+i)\langle\hat\sigma^\dagger_s\rangle]
 +\frac{\gamma_c}{\kappa}\langle\hat \sigma_s^\dagger\hat \sigma_s\rangle
\end{equation}
Now taking the expectation value of Eq. \eqref{8251} along with  \eqref{8255}, we find
\begin{eqnarray}\label{8260}
 \bar n_s&=&\frac{8\eta^2}{\kappa^2}-\frac{4\eta g}{\kappa^2}
 [\langle\hat\sigma_c\rangle+\langle\hat\sigma_{c}^{\dagger}\rangle+\langle\hat\sigma_{c}'\rangle+\langle\hat\sigma_c^{'\dagger}\rangle
 -i(\langle\hat\sigma_c\rangle-\langle\hat\sigma_c^{\dagger}\rangle)+i(\langle\hat\sigma'_{c}\rangle
 -\langle\hat\sigma_c^{'\dagger}\rangle)]\nonumber\\
 &&+\frac{\gamma_c}{\kappa}(\langle\hat \eta_a\rangle+\langle\hat \eta'_a\rangle)
\end{eqnarray}
and applying the fact that $\langle\hat\sigma_c\rangle=\langle\hat\sigma_{c}^{\dagger}\rangle$ and 
$\langle\hat\sigma_{c}'\rangle=\langle\hat\sigma_c^{'\dagger}\rangle$, we have 
\begin{equation}\label{8261}
 \bar n_s=\frac{8\eta^2}{\kappa^2}-\frac{8\eta g}{\kappa^2}
 [\langle\hat\sigma_c\rangle+\langle\hat\sigma_{c}'\rangle]+\frac{\gamma_c}{\kappa}[\langle\hat \eta_a\rangle+\langle\hat \eta'_a\rangle].
\end{equation}
In view of Eqs. \eqref{8213} and  \eqref{8214}, we can write the steady state solutions of the expectation values of $\langle\hat\sigma_{c}'\rangle$ and 
$\langle\hat \eta'_a\rangle$ in the form
\begin{equation}\label{8262}
 \langle \hat \sigma_c'\rangle=\frac{\varepsilon(\gamma_c+\gamma)}{(\gamma_c+\gamma)^2+3\varepsilon^2},
\end{equation}
\begin{equation}\label{8263}
 \langle \hat \eta_a'\rangle=\frac{\varepsilon^2 }{(\gamma_c+\gamma)^2+3\varepsilon^2}.
\end{equation}
Thus upon substituting Eqs. \eqref{8213}, \eqref{8214}, \eqref{8262}, and \eqref{8263} into Eq. \eqref{8261}, we get
\begin{equation}\label{8264}
 \bar n_s=2\left[\frac{4\varepsilon^2}{\kappa\gamma_c}
 -\frac{\varepsilon^2}{\kappa}\left[\frac{(3\gamma_c+4\gamma)}{3\varepsilon^2+(\gamma_c+\gamma)^2}\right]\right].
 \end{equation}
 or
 \begin{equation}\label{8265}
 \bar n_s=2\bar n,
\end{equation}
with $\bar n$ representing the mean photon number for the cavity mode a or b. We see from Eq.\eqref{8265} that
the mean photon number of the superposed cavity modes is twice the mean photon number of cavity mode $a$ or $b$.
\subsection{Quadrature squeezing}
\hspace*{2mm} Here  we wish to calculate the quadrature squeezing of the superposed cavity  modes.
The plus and minus quadrature operators are defined by
\begin{equation}\label{8266}
\hat{c}_{+}=\hat{c}^{\dag}+\hat{c}
\end{equation}
and
\begin{equation}\label{8267}
\hat{c}_{-}=i(\hat{c}^{\dag}-\hat{c}).
\end{equation}
It can be readily established that
\begin{equation}\label{8268}
 [\hat c_-, \hat c_+]=-2i[\hat c, \hat c^\dagger].
\end{equation}
With the aid of Eq. \eqref{8254}, we obtain
\begin{equation}\label{8269}
[\hat c, \hat c^\dagger]=\frac{\gamma_c}{\kappa}(\hat \eta_c+\hat \eta_c'-\hat\eta_a-\hat\eta_a').
\end{equation}
Employing  \eqref{8269} in Eq. \eqref{8268}, we have 
\begin{equation}\label{8270}
 [\hat c_-, \hat c_+]=2i\frac{\gamma_c}{\kappa}[(\hat\eta_a-\hat \eta_c)+(\hat\eta_a'-\hat \eta_c')].
\end{equation}
On account of  Eq. \eqref{8223}, we see that  
\begin{equation}\label{8271}
 \Delta c_{-}\Delta c_{+}\geq \frac{\gamma_c}{\kappa}\left|\langle\hat\eta_a\rangle -
 \langle\hat\eta_c\rangle+\langle\hat\eta_a'\rangle-\langle\hat\eta_c'\rangle\right|.
\end{equation}
Now in view of Eq. \eqref{8215}, one can write 
\begin{equation}\label{8272}
 \langle \hat \eta'_c\rangle=\frac{\varepsilon^2+(\gamma_c+\gamma)^2}{(\gamma_c+\gamma)^2+3\varepsilon^2}.
\end{equation}
Substituting Eqs. \eqref{8214}, \eqref{8215}, \eqref{8263}, and \eqref{8272} into Eq. \eqref{8271}, we readily obtain
\begin{equation}\label{8273}
 \Delta c_{-}\Delta c_{+}\geq f_c(\varepsilon),
\end{equation}
where 
\begin{equation}\label{8274}
 f_c(\varepsilon)=\frac{\gamma_c}{\kappa}\left(\frac{2(\gamma_c+\gamma)^2}{3\varepsilon^2+(\gamma_c+\gamma)^2}\right).
\end{equation} 
\hspace*{2mm} The plus and the minus quadrature variances of the superposed cavity modes are expressible as
\begin{equation}\label{8275}
 (\Delta c_{+})^2=\langle \hat c^{\dagger}\hat c\rangle
 +\langle \hat c\hat c^{\dagger}\rangle + \langle \hat c^{\dagger 2}\rangle + \langle \hat c^{2}\rangle
 -\langle\hat c^{\dagger}\rangle^2-\langle\hat c\rangle^2-2\langle\hat c^\dagger\rangle\langle\hat c\rangle 
\end{equation}
and
 \begin{equation}\label{8276}
 (\Delta c_{-})^2=\langle \hat c^{\dagger}\hat c\rangle
 +\langle \hat c\hat c^{\dagger}\rangle -\langle \hat c^{\dagger 2}\rangle -\langle \hat c^{2}\rangle +\langle\hat c^{\dagger}\rangle^2
 +\langle\hat c\rangle^2-2\langle\hat c^\dagger\rangle\langle\hat c\rangle.
\end{equation}
Employing  Eq. \eqref{8254}, we find 
\begin{equation}\label{8277}
 \langle \hat c\hat c^{\dagger}\rangle=\frac{8\eta^2}{\kappa}-\frac{8\eta g}{\kappa^2}[\langle\sigma_c\rangle+\langle\hat\sigma_c'\rangle]
 +\frac{\gamma_c}{\kappa}[\langle\eta_c\rangle+\langle\eta_c'\rangle],
\end{equation}
\begin{equation}\label{8278}
 \langle \hat c^{2}\rangle=\frac{8\eta^2i}{\kappa^2}-\frac{i8\eta g}{\kappa^2}[\langle\sigma_c\rangle+\langle\hat\sigma_c'\rangle],
\end{equation}
\begin{equation}\label{8279}
 \langle\hat c\rangle^2=\frac{8\eta^2i}{\kappa^2}-\frac{i8\eta g}{\kappa^2}[\langle\sigma_c\rangle+\langle\hat\sigma_c'\rangle]
 +2i\frac{\gamma_c}{\kappa}\langle\hat\sigma_c\rangle\langle\hat\sigma_c'\rangle,
\end{equation}
\begin{equation}\label{8280}
 \langle\hat c^\dagger\rangle\langle\hat c\rangle=\frac{8\eta^2}{\kappa^2}-\frac{8\eta g}{\kappa^2}[\langle\sigma_c\rangle+\langle\hat\sigma_c'\rangle]
 +\frac{\gamma_c}{\kappa}[\langle\hat\sigma_c\rangle^2+\langle\hat\sigma_c'\rangle^2].
\end{equation}
Hence on account of Eqs. \eqref{8261}, \eqref{8277}, \eqref{8278}, \eqref{8279}, and \eqref{8280}, we have 
\begin{equation}\label{8281}
 (\Delta c_{+})^2=\frac{\gamma_c}{\kappa}[\langle\hat \eta_a\rangle+\langle\hat \eta'_a\rangle+\langle\eta_c\rangle+\langle\eta_c'\rangle]
 -2\frac{\gamma_c}{\kappa}[\langle\sigma_c\rangle^2+\langle\hat\sigma_c'\rangle^2],
\end{equation}
\begin{equation}\label{8282}
 (\Delta c_{-})^2=\frac{\gamma_c}{\kappa}[\langle\hat \eta_a\rangle+\langle\hat \eta'_a\rangle+\langle\eta_c\rangle+\langle\eta_c'\rangle]
 -2\frac{\gamma_c}{\kappa}[\langle\sigma_c\rangle^2+\langle\hat\sigma_c'\rangle^2].
\end{equation}
Now using Eqs. \eqref{8213}, \eqref{8214}, \eqref{8215}, \eqref{8262}, \eqref{8263}, and \eqref{8272} in Eqs. \eqref{8281} and \eqref{8282}, we get
\begin{equation}\label{8283}
 (\Delta c_{+})^2=\frac{2\gamma_c}{\kappa}\left[1-\left(\frac{3\varepsilon^4+3\varepsilon^2(\gamma_c+\gamma)^2}
 {(3\varepsilon^2+(\gamma_c+\gamma)^2)^2}\right)\right],
\end{equation}
\begin{equation}\label{8284}
 (\Delta c_{-})^2=\frac{2\gamma_c}{\kappa}\left[1-\left(\frac{3\varepsilon^4+3\varepsilon^2(\gamma_c+\gamma)^2}
 {(3\varepsilon^2+(\gamma_c+\gamma)^2)^2}\right)\right].
\end{equation}
On setting $\varepsilon=0$ in Eqs. \eqref{8283} and \eqref{8284}, the quadrature variance for a vacuum state takes the form
\begin{equation}\label{8285}
 (\Delta c_{\pm})_v^2=2\frac{\gamma_c}{\kappa}.
\end{equation}
\begin{figure}
\centering
 \includegraphics[scale=0.8]{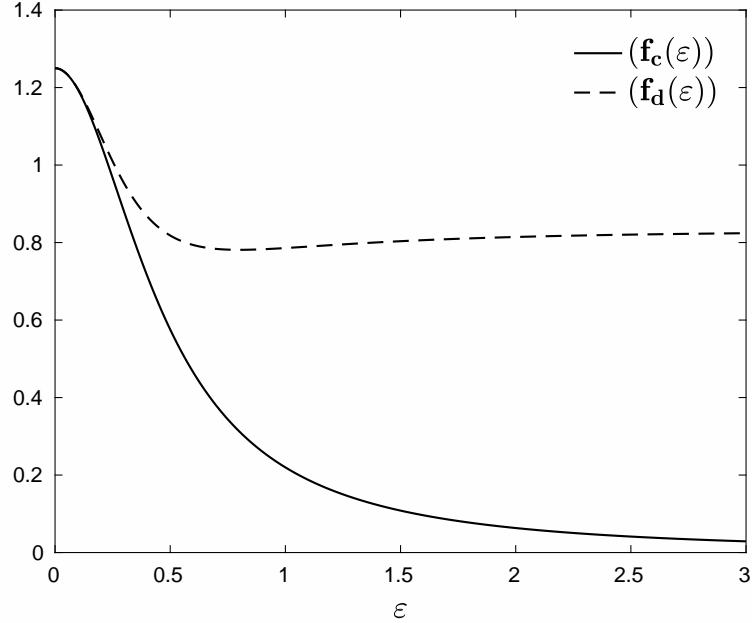}
  \caption{\footnotesize{  Plots of Eqs. \eqref{8274} and \eqref{8287} versus $\varepsilon$ for 
  $\gamma_c=0.5 $, $\kappa=0.8$, and $\gamma=0.3$ }.}
 \label{7}
\end{figure}
Moreover, on account of  Eqs. \eqref{8283} and \eqref{8284}, one can write  
\begin{equation}\label{8286}
 \Delta c_{-}\Delta c_{+}=f_d(\varepsilon),
\end{equation}
where
\begin{equation}\label{8287}
 f_d(\varepsilon)=\frac{2\gamma_c}{\kappa}\left[1-\left(\frac{3\varepsilon^4+3\varepsilon^2(\gamma_c+\gamma)^2}
 {(3\varepsilon^2+(\gamma_c+\gamma)^2)^2}\right)\right].
\end{equation}
Comparison of Eqs. \eqref{8283} and \eqref{8284} with Eq. \eqref{8285} shows that the superposed cavity modes are in a squeezed state and the 
squeezing occurs in both the plus and the minus quadratures.  In addition, 
we note from the plots in Figure \ref{7} that $f_d (\varepsilon) = f_c (\varepsilon)$ 
for values of $\varepsilon$ $0 < \varepsilon < 0.08$ and $f_d (\varepsilon) > f_c (\varepsilon)$ for $\varepsilon > 0.08$.
On the basis of this result, we observe that the quadratures perfectly satisfy the uncertainty relation. This is a similar result found in [23].\\
\hspace*{2mm} We now proceed to calculate the quadrature squeezing of the superposed cavity  modes. To this end, we define the quadrature squeezing of the 
superposed cavity  modes relative to the quadrature variance of the vacuum state by 
\begin{equation}\label{8288}
 S_{sup\pm}=\frac{(\Delta c_{\pm})_v^2-(\Delta c_{\pm})^2}{(\Delta c_{\pm})_v^2}
\end{equation}
Now in view of Eqs. \eqref{8283} and \eqref{8284}, the plus and the minus quadrature squeezing of the superposed cavity modes takes the form
\begin{equation}\label{8289}
 S_{sup+}=1-\left[\frac{6\varepsilon^4+(\gamma_c+\gamma)^2(3\varepsilon^2+(\gamma_c+\gamma)^2)}{(3\varepsilon^2+(\gamma_c+\gamma)^2)^2}\right],
\end{equation}
\begin{equation}\label{8290}
 S_{sup-}=1-\left[\frac{6\varepsilon^4+(\gamma_c+\gamma)^2(3\varepsilon^2+(\gamma_c+\gamma)^2)}{(3\varepsilon^2+(\gamma_c+\gamma)^2)^2}\right].
\end{equation}
Taking into account  Eqs. \eqref{8243} and \eqref{8244}, we can write
\begin{equation}\label{8291}
S_{sup-}= S_{sup+}=\frac{ S_++ S_-}{2}.
\end{equation}
This equation represents the squeezing of the superposed cavity  modes occurs in both quadratures 
and have the same value. The amount of squeezing in each quadrature turns out to be half of the sum of the squeezing in the plus and minus 
quadratures of each cavity mode.  
\section{Conclusion}
\hspace*{2mm} We have found that the variance in the plus and minus quadratures of the cavity mode are below the vacuum state level. However, 
 the product of the uncertainties in the two quadratures perfectly satisfy the uncertainty relation. We have established that 
 the maximum squeezing of the plus and the minus quadratures are 52.08\% and 33.32\% below the vacuum state level, respectively. On the other hand, 
 the squeezing of the superposed cavity modes occurs in both quadratures and have the same value. 
 The amount of squeezing in each quadrature turns out to be half of the sum of the squeezing in the plus and the minus quadratures of each cavity  mode. 
 Furthermore, we have noticed that the  effect of spontaneous emission decreases the mean photon number of the cavity mode 
 but does not affect  the maximum quadrature squeezing. 
 
 \vspace*{5mm}
\noindent
{\bf References}

\vspace*{2mm}
\noindent
[1] D.F. Walls and J.G. Milburn, Quantum Optics (Springer-Verlag, Berlin, 1994).\newline
[2] M. Fox, Quantum optics (Oxford University press Inc., Newyork, 2006).\newline
[3] G.S. Agarwal, Quantum Optics (Cambridge University press, Newyork, 2013).\newline
[4] M.O. Scully and M.S. Zubairy, Quantum Optics (Cambridge University Press, Cambridge, 1997).\newline
[5] P.Meystre and M. Sargent III, Elements of Quantum Optics (Springer-Verlag, Berlin Heidelberg, \hspace*{5.5mm}1990).\newline
[6] J. L. Garrison and R.Y. Chiao, Quantum Optics (Oxford University press, Newyork, 2008).\newline
[7] Z. Ficek and M.R. Wahiddin, Quantum Optics for Beginners (Pan Stanford Publishing, 2014).\newline
[8] A.I. Lvovsky, arXiv:1401.4118v2[quant-ph]28 Jul 2016.\newline
[9] Fesseha Kassahun, Quantum analysis of Light (Kindle Independent Publishing, 2018).\newline
[10] Fesseha Kassahun, arXiv:1508.00342v1[quant-ph] 3 Aug 2015.\newline
[11] M. J. Collett and D. F. Walls, Phys. Rev. A 32, 2887 (1985).\newline
[12] L.X. Zeng  and S. Ying, Phys.Rev. A 40, 12 (1989).\newline
[13] N.A. Ansari, Phys. Rev. A 48, 4686 (1993).\newline
[14] J. Anwar and M.S. Zubairy, Phys. Rev. A 49, 481 (1994).\newline
[15] Fesseha Kassahun, arXiv:1105.1438v3[quant-ph] 25 Sep 2012.\newline
[16] N. Lu, Phys. Rev. A 42, 6756 (1990).\newline
[17] H. Xiong, M. O. Scully, and M. S. Zubairy, Phys. Rev. lett. 023601 (2005).\newline
[18] Eyob Alebachew and K. Fesseha, Opt. Commun. 265, 314 (2006).\newline
[19] Dawit Hailu, Fesseha Kassahun, arXiv:1709.10047v1[quant-ph] 28 Sep 2017.\newline
[20] Fesseha Kassahun, Opt. Commun. 284, 1357 (2011).\newline
[21] Fesseha Kassahun, arXiv:1611.01003v2[quant-ph] 29 Mar 2018.\newline
[22] Fesseha Kassahun, Basic Quantum Mechanics (Elizabeth Printing Press, Addis Ababa, 2016).\newline
[23] Beyene Bashu and Fesseha Kassahun, arXiv:1812.04879v2[quant-ph] 25 May 2019.
\end{document}